\documentclass[a4paper, 11pt]{article}
 
\usepackage[cp1252]{inputenc}
\usepackage[T1]{fontenc}
\usepackage{lmodern}
\usepackage{graphicx}
\usepackage[english, frenchb]{babel}
\usepackage[lined, ruled,linesnumbered]{algorithm2e}
\usepackage{graphicx}
\usepackage{color}
\usepackage{multirow}
\usepackage{epsfig}
\usepackage{amssymb}
\usepackage{amsmath}
\usepackage{amsfonts}
\usepackage{color}
\usepackage{authblk}

\newtheorem{definition}{Definition}

\begin{document}
\title{Improving tag recommendation by folding in more consistency}
\renewcommand\Affilfont{\itshape\small}
\setlength{\affilsep}{1em}
\author[1,2]{Modou Gueye}
\author[1]{Talel Abdessalem}
\author[3]{Hubert Naacke}
\affil[1]{Institut Telecom - Telecom ParisTech\authorcr 46, rue Barrault 75013\authorcr Paris, France\authorcr firstname.lastname@telecom-paristech.fr}
\affil[2]{Universit\'e Cheikh Anta DIOP\authorcr BP. 16432 Fann\authorcr Dakar, S\'en\'egal\authorcr gmodou@ucad.sn}
\affil[3]{LIP6, UPMC Sorbonne Universit\'es - Paris 6\authorcr 4, place Jussieu 75005\authorcr Paris, France\authorcr hubert.naacke@lip6.fr}
\maketitle
\selectlanguage{english}
\begin{abstract}
Tag recommendation is a major aspect of collaborative tagging systems. It aims to recommend tags to a user for tagging an item.
In this paper we present a part of our work in progress which is a novel improvement of recommendations by re-ranking the output of a tag recommender. 
We mine association rules between candidates tags in order to determine a more consistent list of tags to recommend.

Our method is an add-on one which leads to better recommendations as we show in this paper. It is easily parallelizable and morever it may be applied to a lot of tag recommenders.

The experiments we did on five datasets with two kinds of tag recommender demonstrated the efficiency of our method.
\end{abstract}
\section{Introduction}
Social (i.e. collaborative) tagging is the practice of allowing users to annotate content. Users can organize, and search content with annotations called tags. Nowadays the growth in popularity of social media sites has made the area of recommender systems for social tagging systems an active and growing topic of research~\cite{Jaschke:2008:TRS:1487691.1487696,Milicevic:2010:STR:1731524.1731540}. Tag recommenders aim to recommend the most suited tags to a user for tagging an item. They are a salient part of the web 2.0 where applications are user-centred.

In this paper, we propose a novel improvement of tag recommendation. We present $FoldCons$ an add-on method which can fold more consistency in recommendations. Furthermore it is applicable on top of many tag recommenders and is very fast to compute. The main idea behind FoldCons is that the first of recommended tags computed by a tag recommender plays more important role than the rest. We may think that it is the most interesting tag since it has the highest score. Thus FoldCons relies on this first tag to sort the rest in order to achieve better consistency and improvement. Of course, the same reasoning can be used ith the second tag, then the third one and so on.

To validate the efficiency of FoldCons who chose two kinds of tag recommender as candidates. One which proves itself is the pairwise interaction tensor factorization model (PITF) of Rendle and Schmidt-Thieme which wins the task 2 of ECML PKDD Discovery Challenge 2009~\cite{rendle:wsdm10}. Currently one of the best tag recommenders in literature. The other is an adaptation of the network-aware search in online social bookmarking applications of~\cite{Maniu:2012:TEP:2213836.2213926,maniu2011efficient} to tag recommendation, we called STRec~\cite{strec_gan13}. It is a network-based tag recommender which considers the opinions of users' neighbourhood. The experiments we did on five datasets with these two tag recommenders demonstrated the efficiency of FoldCons.

The remainder of this paper is organized as follows. In Section \ref{prelimaries} we present some preliminaries and describe briefly PITF and STRec. Section \ref{FoldCons} details the FoldCons method. In Section \ref{Evaluation}, we present experimentations of our proposal. Finally Section \ref{Related_work} summarizes the related work while Section \ref{Conclusion} concludes the paper.
\section{Preliminaries}
\label{prelimaries}
A folksonomy is a system of classification that allows its users creating and managing tags to annotate and categorize content. It is related to the event of social tagging systems. A folksonomy can be defined as a collection of a set of users $U$, set of tags $T$, set of items $I$, and a ternary relation between them $S \subseteq U \times I \times T$. A tagging triple $(u, i, t) \in S$ means that user $u$ has tagged an item $i$ with the tag $t$. A user can tag an item with one or more distinctive tags from $T$. We assume that a user can tag an item with a given tag at most once.

The interest of a tag $t$ for a given user $u$ and an item $i$ is generally estimated by a score $score(t|u,i)$. Thus the purpose of a tag recommender is to compute the top-\textit{K} highest scoring tags for a post $(u,i)$ what represents its recommendations.
\begin{equation}
    \label{topk_scoring}
    Top(u,i,K) = \overset{K}{\underset{t \in T}{argmax}}~score(t|u,i)
\end{equation}

In the next subsections we describe how PITF and STRec model the scores of tags.
\subsection{Factor Models for Tag Recommendation}
\label{tagrec_func}
Factorization models are known to be among the best performing models. They are a very successful class of models for recommender systems. For tag recommendation they outperform the other approaches like Folkrank and adapted Pagerank~\cite{hotho2006folkrank,symeonidis:recsys08}. We chose the pairwise interaction tensor factorization model (PITF) of Rendle and Schmidt-Thieme in our experimentations due to its efficency~\cite{rendle:wsdm10}. It took the first place of the task 2 of ECML PKDD Discovery Challenge 2009 indeed.
 
PITF proposes to infer pairwise ranking constraints from $S$. The idea is that within a post $(u,i)$, one can assume that a tag $t$ is preferred over another tag $t'$ iff the tagging triple $(u,i,t)$ has been observed and $(u,i,t')$ has not been observed. 
PITF captures the interactions between users and tags as well as between items and tags. Its model equation is given by:
\begin{equation}
    \label{fm_equation}
		score(t|u,i) = \sum_f {\hat{u}_{u,f} \cdot \hat{t}^U_{t,f}} + \sum_f {\hat{i}_{i,f} \cdot \hat{t}^I_{t,f}}
\end{equation}
Where $\hat{U}$, $\hat{I}$, $\hat{T}^U$ and $\hat{T}^I$ are feature matrices capturing the latent interactions. For more information regarding PITF see their paper~\cite{rendle:wsdm10}.
%
\subsection{A Network-based Tag Recommender}
\label{strec_func}
We also used STRec, an adaptation to tag recommendation of the network-aware search in online social bookmarking applications of~\cite{Maniu:2012:TEP:2213836.2213926,maniu2011efficient,ABM-BDA-11}. We chose it as a candidate for network-based tag recommenders. STRec is fast and efficient as presented in~\cite{maniu2011efficient}. It considers that users form an undirected weighted graph $G = (U,E,\sigma)$ (i.e. the social network) where $\sigma$ is a function that associates to each edge $e = (u,v) \in E$ a value in $[0, 1]$, called the proximity between $u$ and $v$. Its model score of a tag $t$ for a post $(u,i)$ is represented by
\begin{equation}
    \label{score_equation}
    score(t|u,i) = h(fr(t|u,i))
\end{equation}
where $fr(t|u,i)$ is the overall frequency of tag $t$ for a user $u$ and item $i$, and $h$ a positive monotone function. In our case we took $h$ as the identity function. They define the overall tag's frequency function $fr(t|u,i)$ as a combination of a user-network-dependent component $sf(t|u,i)$ and an item-dependent one $tf(t,i)$, and as follows:
\begin{equation}
    \label{overall_tag_frequency_equation}
    fr(t|u,i) = \alpha \times tf(t,i) + (1-\alpha) \times sf(t|u,i)
\end{equation}
The former component, $tf(t,i)$, is the frequency of $t$ for item $i$, i.e., the number of times the item was tagged with this tag. The latter component stands for social frequency, an measure that depends on the neighborhood of user $u$. The parameter $\alpha$ allows to tune the relative importance of the social component with respect to tag's frequency.

The scoring model of STRec does not take into account only the close neighborhood of user $u$ (i.e. the other users directly connected to her). But it extends it to deal also with users that are indirectly connected to her, following a natural interpretation that user links (e.g., similarity) are, at least to some extent, transitive.
Thus considering that each neighbour brings her own weight (proximity) to the score of a tag, the measure of tag's social frequency is defined as follows:
\begin{equation}
    \label{social_frequency_equation}
    sf(t|u,i) = \sum_{v \in \left\{U|(v, i, t) \in S\right\}} \sigma(u,v)
\end{equation}

As one may notice, STRec does not regard the use of the tag by the user for tagging items (as for the item with the tag's frequency) but it considers the opinions of the user's neighbours instead.
\section{Folding more consistency in recommendations}
\label{FoldCons}
In this section, we present our add-on method for improving tag recommendation, we called $FoldCons$. It may be used on top of many tag recommenders.\\
Its functioning is to ask recommendations of a given tag recommender about a post $(u,i)$, and improve them before to leave to the user the final top-\textit{K} recommended tags. Therefore it asks for a number of tags greater than $K$. Then it re-ranks them and keeps the $K$ first tags as the final recommendations. The sequel of this section details the FoldCons method besides introducing some defintions.
\begin{definition}
\label{def_tag_lists}
A tag's users list $U(t)$ is the set of users who used the tag $t$. A tag's items list $I(t)$ so is the set of items tagged by the tag $t$.
\end{definition}
\begin{definition}
The pairwise confidence measure, $PCM(t \rightarrow t')$, is defined between to tags $t$ and $t'$. It determines to some extent the interest to use $t'$ in addition to $t$. PCM takes into account both users and items as defined as follows
\begin{equation}
\label{pcm_equation}
	PCM(t \rightarrow t') = \frac{\left\vert{U(t) \cap U(t')}\right\vert}{\left\vert{U(t)}\right\vert} + \frac{\left\vert{I(t) \cap I(t')}\right\vert}{\left\vert{I(t)}\right\vert}
\end{equation}
\end{definition}
The pairwise confidence measure mines association rules between tags from two dimensions: users and items. This allows us to account the frequency of tags' co-occurences both for the user and item of a post. Let us notice there we do not currently weight their contributions in the sum but this is a possibility.
\subsection{FoldCons' functioning}
\label{foldCons_functioning}
FoldCons works simply as an add-on tool which takes in entry tags from a tag recommender sorted by their scores and returns a short list of final recommended tags. Its challenge is to improve the recommendations it received by giving a better top-\textit{K}.

Let us denote by $D$ the list of recommended tags received from a given tag recommender. To simply we consider that $D$ is sorted and its highest scoring tag is $D[1]$. Let us emphasize here we assign more attention to $D[1]$ than the rest of tags in $D$ due to the fact that it is the best choice given by the tag recommender. Therefore we fix it and compute the pairwaise confidence measures of all the tags compared to it. Then we sort $D$ again with the new scoring function for each of its tags $t$
\begin{center}
	$score^{PCM}(t|u,i) = (1 + PCM(D[1] \rightarrow t)) \cdot score(t|u,i)$
\end{center}

Thus we introduce a certain consistency in the recommendations by taking account the tags which appear generally next to the first recommended tag in $D$ from both the user's point of view and the one of the item. This approach improves noticeably the quality of recommendations as shown in our experimentations.

Morever, as PCM at best doubles the initial score of a tag (i.e. $PCM(t \rightarrow t') \in [0,1]$), we can keep in entry only the tags whose scores exceed or equal the half-score of the last tag in the top-\textit{K} of $D$. Indeed the other tags can not change the top-\textit{K}.
\subsection{Still ensuring better recommendations}
\label{adapted_FoldCons}
Some tag recommenders are not both user and item-centred. STRec may be an example. Despite it takes into account the opinions of a user's neighbourhood, it does not consider the user's frequent used tags. In these cases we experimented that the application of FoldCons may slightly fall to improve the recommendations in the user or item-dimension. Thus we adapted FoldCons to these cases. Depending on the recommender we consider the user profile and/or the item one, we define below.

\begin{definition}
\label{def_user_item_profils}
A user profile $T(u)$ is the set of all the tags used by user $u$ to tag items. An item profile $T(i)$ so is the set of tags used to annotate item $i$.
\end{definition}

We determine if FoldCons brings better recommendations by estimating its contribution. We take account of the number of common tags between the top-\textit{K} recommended tags and the item and/or the user profile before and after its computation. The difference represents the contribution of FoldCons.

The recommendations of FoldCons are considered better when its contributions are positive else the list of tags $D$ remains unchanged. This approach ensures, almost in all cases, that the recommendation quality does not decrease after application of FoldCons when recommenders are not both user and item-centred.
\section{Experimentations}
\label{Evaluation}
\subsection{Datasets}
We chose five datasets from four online systems: del.icio.us\footnote{http://www.del.icio.us.com}, Movielens\footnote{http://www.grouplens.org}, Last.fm\footnote{http://www.lastfm.com}, and BibSonomy\footnote{http://www.bibsonomy.org}.

We take the ones of del.icio.us, movielens, and last.fm from \textit{HetRec 2011}~\cite{hetrec2011} and the two other ones from Bibsonomy:  a post-core at level 5 and a one at level 2~\cite{benz2010social,Jaschke:2008:TRS:1487691.1487696}. We call them respectively $Bibson5$ and $dc09$).

$dc09$ is the one of the task 2 of ECML PKDD Discovery Challenge 2009\footnote{http://www.kde.cs.uni-kassel.de/ws/dc09/}. This task was especially intended for methods relying on a graph structure of the training data only. The user, item, and tags of each post in the test data are all contained in the training data's, a post-core at level 2.Let us remaind that a post-core at level $p$ is a subset of a folksonomy with the property, that \textit{each user, tag and item has/occurs in at least $p$ times}.
Table \ref{datasets_tab} presents the caracteristics of these datasets.

\begin{table}[!htp]
  \begin{center}
  	\caption{\label{datasets_tab} Caracteristics of the datasets}
		\begin{tabular}{|c|r|r|r|r|}
			\hline
			dataset & $\left|U\right|$ & $\left|I\right|$ & $\left|T\right|$ & $\left|T(u,i)\right|$ \\ \hline \hline
			Bibson5 & 116 & 361 & 412 & 2,526 \\ \hline
			dc09    & 1,185 & 22,389 & 13,276 & 64,406 \\ \hline
			del.icio.us & 1,867 & 69,226 & 53,388 & 104,799 \\ \hline
			Last.fm & 1,892 & 17,632 & 11,946 & 71,065 \\ \hline
			Movielens & 2,113 & 10,197 & 13,222 & 27,713 \\ \hline
		\end{tabular}
	\end{center}
\end{table}
\subsection{Evaluation Measures and Methodology}
To evaluate our proposal, we used a variant of the leave-one-out hold-out estimation called LeavePostOut~\cite{Jaschke:2008:TRS:1487691.1487696,DBLP:conf/gfkl/MarinhoS07}. In all datasets except $dc09$, we picked randomly and for each user $u$, one item $i$, which he had tagged before. Thus we create a test set and a training one. The task of our recommender was then to predict the tags the user assigned to the item.

Moreover we generate, for each training set, a social network by computing the Dice coefficient of common users' tagged items. Let us notice that we fixed the parameter $\alpha$ of STRec to 0.05 for all experimentations. We kept this value after a calibration over the dataset dc09. What is of course not necessary optimal for all the others. For Tagrec we keep the default parameters given by the authors but with 2,000 iterations\footnote{http://www.informatik.uni-konstanz.de/rendle/software/tag-recommender/}.\\
We used the F1-measure as performance measure.
\subsection{Results}
In this section we present the results of our experimentations on the datasets. On each of these datasets, we run STRec and Tagrec. Then we apply FoldCons on their proposed top-\textit{K} tag. We call respectively by STRec++ and Tagrec++ the application of FoldCons on them. Let us notice that for STRec we specially apply the adapted FoldCons presented in Section \ref{adapted_FoldCons}.
\subsubsection{Contribution of FoldCons}
The tables below show the gains brought by FoldCons when it is applied. We compute the top-\textit{5} to top-\textit{10} recommended tags and their F1-measures.
\begin{table}[h!t]
	\begin{center}
		\caption{\label{DC09} The benefits of FoldCons on dc09}
		\begin{tabular}{|l|c|c|c|c|c|c|}
			\hline
			\#tags & 5 & 6 & 7 & 8 & 9 & 10 \\ \hline \hline
			tagrec & 0.296 & 0.286 & 0.272 & 0.258 & 0.246 & 0.236 \\ \hline
			tagrec++ & 0.301 & 0.290 & 0.279 & 0.265 & 0.251 & 0.241 \\ \hline
			Gain (\%) & \textbf{1.68} & \textbf{1.35} & \textbf{2.74} & \textbf{2.52} & \textbf{2.16} & \textbf{2.23} \\ \hline \hline
			STRec & 0.305 & 0.302 & 0.298 & 0.291 & 0.286 & 0.282 \\ \hline
			STRec++ & 0.309 & 0.312 & 0.306 & 0.297 & 0.292 & 0.285 \\ \hline
			Gain (\%) & \textbf{1.56} & \textbf{3.25} & \textbf{2.55} & \textbf{1.91} & \textbf{2.13} & \textbf{1.18} \\ \hline
		\end{tabular}
	\end{center}
\end{table}
\begin{table}[thp]
	\begin{center}
		\caption{\label{del.icio.us} The benefits of FoldCons on del.icio.us}
		\begin{tabular}{|l|c|c|c|c|c|c|}
			\hline
			\#tags & 5 & 6 & 7 & 8 & 9 & 10 \\ \hline \hline
			tagrec & 0.188 & 0.182 & 0.173 & 0.165 & 0.157 & 0.151 \\ \hline
			tagrec++ & 0.191 & 0.185 & 0.177 & 0.169 & 0.162 & 0.154 \\ \hline
			Gain (\%) & \textbf{1.62} & \textbf{2.09} & \textbf{2.32} & \textbf{2.63} & \textbf{2.67} & \textbf{2.40} \\ \hline \hline
			STRec & 0.103 & 0.105 & 0.107 & 0.107 & 0.108 & 0.109 \\ \hline
			STRec++ & 0.108 & 0.110 & 0.111 & 0.111 & 0.111 & 0.111 \\ \hline
			Gain (\%) & \textbf{5.64} & \textbf{4.73} & \textbf{3.93} & \textbf{3.63} & \textbf{3.02} & \textbf{2.29} \\ \hline
		\end{tabular}
	\end{center}
\end{table}
\begin{table}[h!t]
	\begin{center}
		\caption{\label{lastfm} The benefits of FoldCons on last.fm}
		\begin{tabular}{|l|c|c|c|c|c|c|}
			\hline
			\#tags & 5 & 6 & 7 & 8 & 9 & 10 \\ \hline \hline
			tagrec & 0.328 & 0.309 & 0.290 & 0.272 & 0.256 & 0.242 \\ \hline
			tagrec++ & 0.333 & 0.314 & 0.295 & 0.278 & 0.261 & 0.246 \\ \hline
			Gain (\%) & \textbf{1.61} & \textbf{1.86} & \textbf{1.95} & \textbf{1.95} & \textbf{1.81} & \textbf{1.76} \\ \hline \hline
			STRec & 0.274 & 0.260 & 0.246 & 0.235 & 0.224 & 0.215 \\ \hline
			STRec++ & 0.277 & 0.262 & 0.248 & 0.237 & 0.225 & 0.216 \\ \hline
			Gain (\%) & \textbf{1.15} & \textbf{0.80} & \textbf{0.99} & \textbf{0.87} & \textbf{0.80} & \textbf{0.58} \\ \hline
		\end{tabular}
	\end{center}
\end{table}
\begin{table}[h!t]
	\begin{center}
		\caption{\label{bibsonomy} The benefits of FoldCons on bibson5}
		\begin{tabular}{|l|c|c|c|c|c|c|}
			\hline
			\#tags & 5 & 6 & 7 & 8 & 9 & 10 \\ \hline \hline
			tagrec & 0.449 & 0.426 & 0.409 & 0.390 & 0.371 & 0.353 \\ \hline
			tagrec++ & 0.450 & 0.429 & 0.412 & 0.390 & 0.373 & 0.357 \\ \hline
			Gain (\%) & \textbf{0.30} & \textbf{0.58} & \textbf{0.67} & \textbf{0.05} & \textbf{0.65} & \textbf{1.07} \\ \hline \hline
			STRec & 0.389 & 0.373 & 0.360 & 0.349 & 0.340 & 0.334 \\ \hline
			STRec++ & 0.397 & 0.379 & 0.364 & 0.352 & 0.343 & 0.337 \\ \hline
			Gain (\%) & \textbf{2.00} & \textbf{1.80} & \textbf{0.97} & \textbf{0.97} & \textbf{0.88} & \textbf{0.94} \\ \hline
		\end{tabular}
	\end{center}
\end{table}
\begin{table}[h!t]
	\begin{center}
		\caption{\label{movielens} The benefits of FoldCons on movielens}
		\begin{tabular}{|l|c|c|c|c|c|c|}
			\hline
			\#tags & 5 & 6 & 7 & 8 & 9 & 10 \\ \hline \hline
			tagrec & 0.163 & 0.148 & 0.135 & 0.124 & 0.115 & 0.108 \\ \hline
			tagrec++ & 0.164 & 0.148 & 0.136 & 0.125 & 0.116 & 0.108 \\ \hline
			Gain (\%) & \textbf{0.70} & \textbf{0.34} & \textbf{0.70} & \textbf{0.67} & \textbf{0.69} & \textbf{0.67} \\ \hline \hline
			STRec & 0.146 & 0.138 & 0.131 & 0.127 & 0.122 & 0.119 \\ \hline
			STRec++ & 0.148 & 0.140 & 0.133 & 0.128 & 0.124 & 0.120 \\ \hline
			Gain (\%) & \textbf{1.57} & \textbf{1.46} & \textbf{1.56} & \textbf{1.08} & \textbf{0.98} & \textbf{1.06} \\ \hline
		\end{tabular}
	\end{center}
\end{table}
As one can notice, the efficiency of FoldCons is not the same on all the five datasets. Indeed FoldCons clearly improves the recommendations for $del.icio.us$ and $dc09$ datasets.
But it struggles to improve the recommendations on the other three datasets.\\ 

We can explain this by the fact that the pairwise confidence measure depends only on the tag's users list $U(t)$ and the tag's items one $I(t)$ (see Equation \ref{pcm_equation}). Thus if these lists are short, the probability to retrieve co-occurences between two lists becomes low. Therefore the PCM's value may be small.\\
Let us note that the size of the tag's users list (resp. the tag's items one) may be related to the average number of posts of users (resp. items). Indeed one can assume that more posts a user has, more distinct tags she might use. Therefore when the user's average number of posts is great, we expect more impact when applying PCM. Table \ref{connection_nbPosts_efficiency} below confirmes this analysis. It shows that for the datasets where the average number of posts of users is greater than 50 (i.e. $del.icio.us$ and $dc09$), the gain led by FoldCons exceeds 2\%. But this gain remains slight where this average number of posts is small (e.g., lesser than 40). This is a weak point of the PCM method. 
\begin{table}[h!t]
	\begin{center}
		\caption{\label{connection_nbPosts_efficiency} Connection between the users' average number of posts and the efficiency of FoldCons}
		\begin{tabular}{|l|r|r|r|}
			\hline
			\multirow{2}{*}{Dataset} & \multirow{2}{*}{\#posts/user} & \multicolumn{2}{c|}{Average gain (\%) on} \\
			\cline{3-4} &            & STRec        & Tagrec \\ \hline \hline
			del.icio.us   & 56.13      & 3.87         & 2.29   \\ \hline
			dc09        & 54.35      & 2.10         & 2.11   \\ \hline \hline
			lastfm      & 37.56      & 0.86         & 1.82   \\ \hline
			bibsonomy   & 21.77      & 1.26         & 0.55   \\ \hline
			movielens   & 13.11      & 1.29         & 0.63   \\ \hline
		\end{tabular}
	\end{center}
\end{table}
\subsubsection{Why assigning more attention to the first tag ?}
In Section \ref{foldCons_functioning}, we emphasized that we assign more attention to the first tag than the rest of tags in the list we received. We assumed that the list is sorted and its highest scoring tag is the first tag.\\
We explained this position by the fact that the first tag is the best choice given by the tag recommender. Therefore we fix it and compute the pairwaise confidence measures of all the tags compared to it.\\

In this section, we experimented the pertinence of this position. We compared the gains brought by FoldCons when we use the first tag as reference tag, or the second and so on. We just show there the results of experimentations for the top-\textit{5} tags to recommend because the results do not change a lot for top-\textit{6} to top-\textit{10} tags to recommend.
These experimentations support more our position of giving more attention to the first tag. Indeed, almost in 80\% of cases, this choice gives the best improvements. Tables \ref{comparison_refTag_strec} and \ref{comparison_refTag_tagrec} present the gains obtained for each of the four first tags when we use it as the reference one.
\begin{table}[h!t]
	\begin{center}
		\caption{\label{comparison_refTag_strec} Comparison of gains for different reference tags with STRec}
		\begin{tabular}{|l|r|r|r|r|}
			\hline
			\multirow{2}{*}{Dataset} & \multicolumn{4}{c|}{Reference tag} \\
			\cline{2-5} & $1^{st}$ & $2^{nd}$ & $3^{th}$ & $4^{th}$ \\ \hline \hline
			bibsonomy & 1.776 & \textbf{2.388} & 1.212 & 0.873 \\ \hline
			movielens & \textbf{3.005} & 2.169 & 1.76 & 0.955 \\ \hline
			del.icio.us & \textbf{4.091} & 3.857 & 3.537 & 1.93 \\ \hline
			lastfm & \textbf{0.958} & 0.75 & 0.601 & 0.473 \\ \hline
			dc09 & 2.344 & \textbf{2.945} & 2.295 & 1.966 \\ \hline
		\end{tabular}
	\end{center}
\end{table}
\begin{table}[h!t]
	\begin{center}
		\caption{\label{comparison_refTag_tagrec} Comparison of gains for different reference tags with Tagrec}
		\begin{tabular}{|l|r|r|r|r|}
			\hline
			\multirow{2}{*}{Dataset} & \multicolumn{4}{c|}{Reference tag} \\
			\cline{2-5} & $1^{st}$ & $2^{nd}$ & $3^{th}$ & $4^{th}$ \\ \hline \hline
			bibsonomy & 0.285 & 0.224 & 0.0 & \textbf{0.741} \\ \hline
			movielens & \textbf{0.697} & 0.172 & 0.081 & 0.119 \\ \hline
			del.icio.us & \textbf{2.766} & 2.365 & 2.129 & 1.649 \\ \hline
			lastfm & \textbf{1.615} & 1.342 & 1.148 & 1.001 \\ \hline
			dc09 & \textbf{1.686} & 0.296 & 1.561 & 1.049 \\ \hline
		\end{tabular}
	\end{center}
\end{table}
\newpage
\subsubsection{Using several tags as reference}
One can wonder why not using several tags as reference instead of the first tag only. The results in Tables \ref{comparison_nBrefTag_strec} and \ref{comparison_nBrefTag_tagrec} give an answer (for recommendations of 5 tags). They point out the ability of this approach to pass the use of one tag as reference. But when we compare these two tables, it seems difficult to define the ideal number of tags to use as reference. That is why we limit ourselves to the first tag only.
\begin{table}[h!t]
	\begin{center}
		\caption{\label{comparison_nBrefTag_strec} Comparison of gains with several tags as reference with STRec}
		\begin{tabular}{|l|r|r|r|r|}
			\hline
			\multirow{2}{*}{Dataset} & \multicolumn{4}{c|}{Using the \# first tags} \\
			\cline{2-5} & 1 & 2 & 3 & 4 \\ \hline \hline
			bibsonomy & 1.776 & 1.805 & \textbf{2.225} & 1.354 \\ \hline
			movielens & 3.005 & 2.963 & \textbf{3.244} & 1.969 \\ \hline
			del.icio.us & 4.091 & 3.705 & \textbf{4.443} & 3.12 \\ \hline
			lastfm & 0.958 & 0.93 & \textbf{1.026} & 0.698 \\ \hline
			dc09 & 2.344 & 2.737 & \textbf{2.8} & 2.585 \\ \hline
		\end{tabular}
	\end{center}
\end{table}
\begin{table}[h!t]
	\begin{center}
		\caption{\label{comparison_nBrefTag_tagrec} Comparison of gains with several tags as reference with Tagrec}
		\begin{tabular}{|l|r|r|r|r|}
			\hline
			\multirow{2}{*}{Dataset} & \multicolumn{4}{c|}{Using the \# first tags} \\
			\cline{2-5} & 1 & 2 & 3 & 4 \\ \hline \hline
			bibsonomy & 0.285 & \textbf{0.756} & 0.0 & 0.276 \\ \hline
			movielens & 0.697 & 0.857 & 0.697 & \textbf{1.145} \\ \hline
			del.icio.us & 2.766 & 2.819 & 3.719 & \textbf{3.972} \\ \hline
			lastfm & 1.615 & \textbf{1.623} & 1.181 & 1.248 \\ \hline
			dc09 & 1.686 & 1.611 & \textbf{1.923} & 1.416 \\ \hline
		\end{tabular}
	\end{center}
\end{table}
\subsubsection{Performance of each component of PCM}
The definition of PCM presented in Section \ref{FoldCons} by Equation \ref{pcm_equation} points out it mines association rules between tags from two dimensions: users and items. What allows it to account the frequency of tags' co-occurences both for the user and item of a post.\\

We would know there the performance of each of its two components (i.e. dimensions) comparing to their linear combination used until there.\\
Tables \ref{combination_strec} and \ref{combination_tagrec} present the results in term of improvement (in percentage) brought by the components.
\begin{table}[h!t]
	\begin{center}
		\caption{\label{combination_strec} Performance of PCM components with STRec}
		\begin{tabular}{|l|r|r|r|}
			\hline
			\multirow{2}{*}{Dataset} & \multicolumn{3}{c|}{Used dimension of PCM} \\
			\cline{2-4} & Item & User & Both \\ \hline \hline
			bibsonomy & \textbf{2.388} & 2.006 & 1.783 \\ \hline
			movielens & 0.253 & \textbf{3.005} & 1.858 \\ \hline
			del.icio.us & 3.757 & 3.915 & \textbf{4.091} \\ \hline
			lastfm & 0.864 & 0.802 & \textbf{0.958} \\ \hline
			dc09 & 2.182 & \textbf{2.945} & 2.295 \\ \hline
		\end{tabular}
	\end{center}
\end{table}
\begin{table}[h!t]
	\begin{center}
		\caption{\label{combination_tagrec} Performance of PCM components with Tagrec}
		\begin{tabular}{|l|r|r|r|}
			\hline
			\multirow{2}{*}{Dataset} & \multicolumn{3}{c|}{Used dimension of PCM} \\
			\cline{2-4} & Item & User & Both  \\ \hline \hline
			bibsonomy & 0.506 & 0.0 & \textbf{0.741} \\ \hline
			movielens & 0.238 & 0.0 & \textbf{0.697} \\ \hline
			del.icio.us & \textbf{2.766} & 1.091 & 1.983 \\ \hline
			lastfm & 0.517 & 0.378 & \textbf{1.615} \\ \hline
			dc09 & 0.959 & 0.765 & \textbf{1.686} \\ \hline
		\end{tabular}
	\end{center}
\end{table}
Despite that is not always the case, the no-weighted combination of the two components proves to be better than the use of one of these component separately. This confirm our initial choice. Morever we think that it would be more obvious when using weighted linear combinations.
\section{Related Work}
\label{Related_work}
Finding suited tags to put in the same recommendations is an important point for tag recommendation. Many approaches and methods can be used to achieve this point.
We can cite the work of Lipczak which focused on content-based tag recommenders~\cite{lipczak2008}. His approach consists in extracting basic tags from the content of items (e.g. the item title), then extending the set of potential recommendations by related tags proposed by a lexicon based on co-occurrences of tags within item's posts. He solves these co-occurences based on association rules mining.

Wang et al. did a similar work enough but first applied a TF-IDF algorithm on the description of the item content, in order to extract keywords of the item~\cite{Wang:ecml2009}. Based on the top keywords, they utilize association rules from history records in order to find the most probable tags to recommend. In addition, if the item has been tagged before by other users or the user has tagged other items before, then history information is also exploited to find the most appropriate recommendations.\\

Many others works could be cited. However due to the length of the paper we can not cite them and furthermore they are generally closed approaches. They are for the most part content-dependent. What is not the case of FoldCons which mines association rules directly on a primary list of candidate tags, then sort this list again. Our experimentations showed the effectiveness of this method.
\section{Conclusion}
\label{Conclusion}
We proposed an add-on method to improve the recommendations of tag recommender. We mine association rules on top of their recommendations, then we sort them again thanks to their confidence scores compared to the first tag. Thus we introduce a certain consistency inside the recommendations by taking account of tags which appear generally next to the first candidate tag in the initial recommended list.

This method improves up to 5\% the recommendations as shown by our experimentations when the users have not in average a small number of posts.
\section{Acknowledgements}
We would like to thank Steffen Rendle for the software \textit{tagrec} and also the Knowledge and Data Engineering Group of University of Kassel especially for the Benchmark Folksonomy Data (BibSonomy, version of April 30th, 2007).
\small

\end{document}